\documentclass{elsart}
\usepackage{natbib}

\journal{Journal of Theoretical Biology}

\begin{document}
\begin{frontmatter}
\title {Plant-Mycorrhiza Percent Infection as Evidence of Coupled Metabolism}
\author{Reginal D. Smith}
\address {Bouchee-Franklin Research Institute, PO Box 10051, Rochester, NY 14610}
\date{January 25, 2009}
\ead {rsmith@sloan.mit.edu}


\begin{abstract}
A common feature of mycorrhizal observation is the growth of the
infection on the plant root as a percent of the infected root or
root tip length. Often, this is measured as a logistic curve with an
eventual, though usually transient, plateau. It is shown in this
paper that the periods of stable percent infection in the
mycorrhizal growth cycle correspond to periods where both the plant
and mycorrhiza growth rates and likely metabolism are tightly
coupled.
\end{abstract}
\maketitle
\end{frontmatter}

\section{Introduction to Mycorrhiza}
Among the many types of symbioses, mutualisms are often one of the
most interesting where organisms engage in mutually beneficial
relationships in order to enhance the survival and adaptability of
each. Among mutualisms, one of the most studied yet still surprising
is the mycorrhizal fungus relationship with plants. Mycorrhizae are
fungi which are adapted to live in a mutualistic association with
plants by growing on the plant roots and providing nutrients,
usually phosphorus, nitrogen, and heavy metals such as copper, in
return for carbohydrates for growth. There are two main types:
ectomycorrhiza whose hyphae cover the external part of the plant
root and endomycorrhiza whose hyphae penetrate the root cells and
deliver and receive nutrients directly. For the fungus, it is an
obligate relationship since mycorrhiza never exist in nature outside
of the mutualistic relationship and are even difficult to grow and
culture in laboratory settings. For the plant, the relationship is
facultative for most environments even though 80\% of vascular plant
species can form symbioses with arbuscular mycorrhiza
\citep{mycorrhiza3}. Mycorrhiza coevolved with the emergence of
vascular plants on land and are believed to have first appeared more
than 400 million years ago \citep{mycorrhiza2}.

In particular, this paper focuses on the most common and most
studied mycorrhiza are the endomycorrhizal arbuscular mycorrhiza
(AM). AM fungi growth on plants is usually characterized by percent
infection of the root. In this measurement, the roots are cleaned,
often with potassium hydroxide in an autoclave, and then stained by
trypan blue, chlorazol black or another appropriate stain to reveal
the AM hypahe on the roots. Using a grid intersection method on a
grid in a petri dish, one can estimate the percent of the root
covered by (``infected with") mycorrhiza and determine the percent
infection. In the course of normal growth, the percent of the root
length infected by AM hyphae follows a logistic growth curve with a
phase of rapid infection after the initiation of the infection by
spores in the soil, to a carrying capacity which it maintains,
absent external influences \citep{mycorrhiza1,logistic1}. Percent
infection growth is not always so clean though, because carrying
capacity can change over time due to factors affecting root growth
or environmental factors. While this paper does not seek to
determine the exact biochemical or ecological mechanisms leading to
this growth curve, there is a possible connection with the
metabolism of both the plant and mycorrhiza that will be explored.

\section{Mathematical Preliminaries}

Investigating the percent infection curve, we will concentrate on
the carrying capacity steady state. Where $M$ is the length of
intercellular hyphae on the average root in an infected plant and
$P$ is the average root length, percent infection is defined by

\begin{equation}
\label{eq1}
C = \frac{M}{P}
\end{equation}

The two stable fixed points of the mycorrhizal percent infection are
0\% and $C = C_{m}$, the maximum percent infection under given
conditions. Obviously at this point

\begin{equation}
\label{eq2}
\frac{dC}{dt} = 0
\end{equation}

Representing equation \ref{eq2} in terms of M and P, we get

\begin{equation}
\label{deriv1} \frac{P\frac{dM}{dt}-M\frac{dP}{dt}}{P^2} = 0
\end{equation}

and eventually

\begin{equation}
\label{deriv2} \frac{dM}{dP}= \frac{M}{P}
\end{equation}

or

\begin{equation}
\label{deriv3} \frac{dM}{M}= \frac{dP}{P}
\end{equation}

Equation \ref{deriv3} essentially shows that at carrying capacity,
or any other steady state of percent infection which may differ over
time, the mycorrhiza percentage growth and the plant root percentage
growth are equal. Therefore, though the different rates of growth
for the two organisms differ, their relative growth is identical.
This exercise may seem like a mathematical triviality but raises a
tantalizing question, what does it mean when two organisms thus
intertwined have coordinated growth patterns?

\section{Root and Mycorrhiza Growth: Exponential Growth or Allometric Scaling?}

In order to dig deeper into this phenomenon, we need to talk more
about the growth of the individual organisms and how the percent
infection is a reflection of their dependency
\citep{percentinfection}. Here we are looking at root growth and the
growth of the mycorrhizal hyphae.

There have not been extensive studies done on the relationships of
root growth and root length and almost none on the intrinsic growth
formula of mycorrhizal hyphae. Given lack of firm footing in these
relationships, here it will be shown that two common growth models
often assumed, the exponential growth model and the allometric
scaling model, both lead to the same results without assuming
\emph{a priori} the value of the growth constants or allometric
scaling exponents.

\section{Derivation of Metabolic Coupling}

In \citep{logistic4, percentinfection, plantscale4} exponential
growth models for root length are assumed. In the case of
\citep{transfer,percentinfection, plantscale4} an explicit equation
of the form

\begin{equation}
\frac{dP}{dt} = rP
\end{equation}

and
\begin{equation}
P = P_{0}e^{rt}
\end{equation}

is assumed where $r$ is the growth rate and is calculated in the
paper from laboratory measurements of root lengths in
\citep{percentinfection} and theoretical considerations in
\citep{transfer,plantscale4}. In \citep{logistic4} a growth curve of
root length over time fitting an exponential is shown though no
equation is explicitly derived explaining this fact. Since most
papers on mycorrhizal infection focus on percent infection as a
measurement, few papers actually try to account for the actual
length of mycorrhizal hyphae or their growth relationship.

Given these assumptions, and assuming $\alpha_{1}$ and $\alpha_{2}$
are the growth rates of the mycorrhizal hyphae and plant roots
respectively

\begin{equation}
\label{coupled6} \frac{dM}{dP} = \frac{\alpha_{1}M} {\alpha_{2}P} =
\frac{M}{P}
\end{equation}

so
\begin{equation}
\label{coupled7}
\frac{\alpha_{1}} {\alpha_{2}} = 1
\end{equation}

\begin{equation}
\label{coupled8}
\alpha_{1} =\alpha_{2}
\end{equation}

where the growth rate constants have a fixed relationship of
equality. This demonstrates that at a steady state of percent
infection, the growth rates of both the intercellular mycorrhizal
hyphae and plant roots are equal. Given the nature of resource
exchange, it is likely that some sort of stabilized feedback between
the resource exchange and metabolism of both organisms has been
reached.

Another model of the growth of structures in organisms is allometric
scaling. Allometric scaling relationships between metabolisms and
growth rates, organ sizes, or life spans are one of the most
ubiquitous relationships among all biological organisms at all
scales. The presence of such universal laws, which usually postulate
a power law relationship between metabolism and organism size, have
inspired countless explanations. The first mathematical models by
P\"{u}tter and von Bertalanffy \citep{plantscale10} still hold
though recently the most well-known and controversial theory is the
fractal origin theory of West, Brown and Enquist \citep{westbig}.
The relationship is usually of the form

\begin{equation}
\label{metabeq} M = C_{0}Y^{\gamma}
\end{equation}

where $M$ can be population density, life span, metabolic rate or
another similar characteristic. $C_{0}$ is the normalization
constant, $Y$ is a morphological measurement (such as body size),
and $\gamma$ is the power law scaling exponent, which has often been
measured, though far from universally, as 3/4. It is not surprising
therefore that this relationship has been extrapolated to botany
with measurements of life span vs. body mass, metabolic scaling vs.
stem diameter or other measurements, growth rates of roots or stems
vs. length, etc. \citep{plantscale9, plantscale7, plantscale5,
plantscale2, plantscale3,plantscale, plantscale6, plantscale8}.
Though there is still a lively argument over what the standard
scaling exponent, if any, exists between these many different
measurements, it is agreed that allometric scaling applies to plants
just as any other organisms.

For this paper, what is of most interest is allometric scaling with
the length of the root. \citep{plantscale6}, postulates root growth
scales isometrically with the length and the square of root diameter
which would be necessary to keep a constant root density for
increasing mass. This growth, like the metabolic scaling with body
size, is directly proportional to the metabolic rate which is what
scales with the size of morphological structures. Root length, like
many morphological quantities such as body size or stem length, have
a growth rate that scales as a power law with their current size.
Therefore, we can hypothesize the growth of a root of length $P$ as

\begin{equation}
\label{rootgrow} \frac{dP}{dt} = C_{1}P^{\gamma_{1}}
\end{equation}

In this equation, $\gamma_{1}$ is the scaling constant for the root
growth over time. Mass growth has often been tied to metabolic rate
and thus body mass \citep{westbig,plantscale10}. Allometric scaling
studies have not yet been carried out on mycorrhizal fungi.
Therefore, it is pure a postulate that the growth of hyphae or
mycelial structures follows a similar relationship, though it is
likely given the trends across various kingdoms of life. However,
given that fungi are heterotrophs, the scaling constant could be
different from plants.

Assume that the growth of the mycorrhiza also exhibit allometric
scaling of the form

\begin{equation}
\label{mycogrow}
\frac{dM}{dt} = C_{2}M^{\gamma_{2}}
\end{equation}

We do not need to assume or calculate as in other papers the value
of either scaling constant $\gamma_{1}$ or $\gamma_{2}$.

Let us now return to the earlier question of how the mycorrhiza and
plant can match their percentage growth rates at the carrying
capacity for percent infection. Given equation \ref{deriv2} and
equations \ref{rootgrow} and \ref{mycogrow} we can show

\begin{equation}
\label{coupled1} \frac{dM}{dP}
=\frac{C_{2}M^{\gamma_{2}}}{C_{1}P^{\gamma_{1}}} = \frac{M}{P}
\end{equation}

Next, we consider with each proportional growth $b$ of the plant
root, the mycorrhiza must also grow proportionally to keep the
percent infection constant so for example

\begin{equation}
\label{coupled1b}
\frac{bM}{bP} =  \frac{M}{P}
\end{equation}

however given equation \ref{coupled1} we also have the growth terms
where

\begin{equation}
\label{coupled2}
\frac{C_{2}(bM)^{\gamma_{2}}}{C_{1}(bP)^{\gamma_{1}}}
\end{equation}

and

\begin{equation}
\label{coupled3}
\frac{b^{\gamma_{2}}C_{2}(M)^{\gamma_{2}}}{b^{\gamma_{1}}C_{1}(P)^{\gamma_{1}}}
= \frac{M}{P}
\end{equation}

where given the equality demonstrated earlier in equation
\ref{coupled1}
\begin{equation}
\label{coupled4} \frac{b^{\gamma_{2}}}{b^{\gamma_{1}}}
=b^{\gamma_{2}-\gamma_{1}} = 1
\end{equation}

we finally have

\begin{equation}
\label{coupled5} \gamma_{1} =\gamma_{2}
\end{equation}

Therefore, even under a model of growth via allometric scaling, the
steady state of percent infection still implies a strong coupling of
the metabolism of the mycorrhizal hyphae and the plant roots. As
stated earlier, allometric scaling as a model for the growth of
these structures is still a premature hypothesis but was shown in
order to emphasize the likelihood of the metabolic coupling this
paper postulates.

\section{Root turnover and arbuscule cycle}

The growth of both the mycorrhiza and root are continuous, however,
there is also turnover of older structures. For the mycorrhiza, this
is dominated by a cycle within the plant cells that can last for
several days \citep{transfer, review} but ends with the degeneration
of the arbuscule and the release of its cytoplasm into the plant
cell. For the root, root turnover plays a similar role. Therefore
the growth rates represented in the equations should be considered
net growth rates which incorporate additional roots or hyphae minus
the root or hyphae turnover.

Rapid root turnover usually can consume a sizable portion of plant
carbon but some studies \citep{turnover2} demonstrate that with
mycorrhizal infection, root turnover slows possibly because of a
large carbon allocation diverted to the mycorrhiza instead of root
structures. Mycorrhizal infection can also be reduced under
conditions of forced high root turnover such as tilled soil
\citep{tilled}. This allocation of carbon away from the plant roots
to the mycorrhiza could be part of the metabolic coupling that the
mycorrhiza undergoes during the infection.

In addition, though the mycorrhizal arbuscules often disintegrate
within the cell releasing their cytoplasm, it is unlikely that this
is a major mechanism of phosphorus transfer to the plant
\citep{transfer}. However, the autolysis of the arbuscules and the
reduction of activity among the intercellular hypahe increases with
the age of the plant \citep{age1}. Therefore, the growth of the
mycorrhizal hyphae may not be just to keep up with the plant root
growth but also to provide fresh structures for the symbiosis to
replace degenerated ones.

Given continuous turnover, a major aim of the mycorrhiza's growth in
steady state may be to ensure the continued bidrectional flow of
nutrients. If the mycorrhiza did not expand colonization at a
minimum of the rate of hyphae turnover, the infection would
disappear over time and decrease in importance. In addition,
matching root growth would allow the mycorrhiza to continue to
receive carbohydrates by keeping up its steady flow of phosphate
compounds.

\section{Implications based on phosphate/carbon exchange}

The exchange of phosphate from AM to the plant and the reciprocal
exchange of carbohydrates from plant to AM has always been an area
of frequent research, but has recently been aided by tools in
molecular biology and genetics to examine the expression of genes in
both mycorrhiza and plants in response to the symbiosis; see
\citep{carbon3,carbon2,carbon4,phospho3,phospho2,phospho1,carbon1}.
Although the exact nature of the pathway which facilitates the
exchange of phosphates, transported through the fungus as
polyphosphate chains, and carbohydrates, mainly in the form of
hexose, has not yet been fully worked out, there are many clues to
follow. For example, in \citep{carbon1} it is demonstrated that the
hexose provided to the mycorrhiza by the plant is regulated by the
symbiosis and cannot be increased by simply adding more hexose to
roots. In addition, there is the common result of reduced AM
colonization in the presence of increased phosphate levels to the
plant \citep{logistic3,logistic2}.

From this and other clues, there is a method to test the assumption
of this paper in a carefully controlled experiment. First, one could
set up a soil region separated by mesh fine enough for mycorrhiza
but too large for roots similar to \citep{meshexp,meshexp2}. The
mycorrhiza accessible region will have radioactive $^{33}P$ added to
the soil and then simultaneously the plant will be provided with
radioactive labeled hexose such as in \citep{carbonlabel, carbon2}.
By measuring the percent infection for the mycorrhiza on the plant
roots and simultaneously measuring the rate of transfer of labeled
hexose to the mycorrhiza and vice versa for labeled phosphorus,
relative rates of the nutrient exchange can be determined. According
to this analysis, at the plateau for percent infection, the relative
nutrient exchanges should maintain a constant ratio even if they
increase over time with the coupled growth of the plant and
mycorrhiza.

\section{Discussion}

The author here is the first to admit, no experimental verification
of coupled metabolism or growth rates between mycorrhiza have been
performed. There is some indirect evidence such as the percent
infection curves, reduced colonization in the presence of increased
phosphorus to the plant, and controlled supply of hexose by the
plant even in the face of a surplus. Even if the paper's thesis is
correct, this knowledge does not yet allow us to directly calculate
the maximum percent infection or make any reliable predictions.
However, it can help give this steady state a deeper biological
meaning.

Results from other experiments seem to indicate that the
mycorrhiza's growth rate is heavily influenced not just by its size
but by the nutrition provided by the plant in the symbiosis. When
compounds that the mycorrhiza usually supply, such as phosphorus,
become more plentiful either in fertilizer or by being added to
leaves the percent infection diminishes. With less needs the plant
likely provides less nutrients to the mycorrhiza forcing it to
reduce its relative growth in order to match the plants metabolism
with lower level of carbohydrate inputs.

The result in this paper is interesting because it shows how
hundreds of millions of years of coevolution have caused two species
in symbiosis to evolve to the point where one adapts to the
metabolism of the other to survive. How widespread this effect is in
biology is an open question and experimental verification is
necessary to fully confirm the nature and extent of the coupling
between plants and mycorrhiza.

\end{document}